\documentclass[12pt, a4paper]{article}
\usepackage[utf8]{inputenc}
\usepackage{hyperref}
\usepackage{geometry}
\usepackage{graphicx}
\usepackage{subcaption}
\usepackage[numbers]{natbib}
\usepackage{amssymb}                                                                                                                                                                                                                                                                                                                                                                                                                                                                                                                                                                                                                                                                                                                                                                                                                                                                                                                                                                                                                                                                                                                                                                                                                                                                                                                                                                                                                                                                                                                                                                                                                                                                                                                                                                                                                                                                                                                                                                                                                                                                                                                                                                                                                                                                                                                                                                                                                                                                                                                                                                                                                                                                                                                                                                                                                                                                                                                                                                                                                                                                                                                                                                                                                                                                                                                                                                                                                                                                                                                                                                                                                                                                                                                                                                                                                                                                                                                                                                                                                                                                                                                                                                                                                                          
\usepackage{amsmath}
\usepackage{slashed}
\usepackage{anyfontsize}
\usepackage{bm}
\usepackage{slashed}
\usepackage{multirow}
\usepackage{float}
\usepackage[font=small]{caption}
\usepackage{listings}
\newgeometry{lmargin=1.1in, rmargin=1.1in, tmargin=1in, bmargin=1in}
\title{The complete metric study of effective Dirac algebra}
\author{B.T.T.Wong\footnote{CERN, u3500478@connect.hku.hk}}
\date{}

\begin{document}

\maketitle
\begin{abstract}
Following our work from the previous paper about the study of effective Dirac algebra and the metric of the simple, special case of relativistic hydrogen atom, this paper gives the complete metric study defined by the effective Dirac algebra in the Dirac and Weyl presentation, showing that relativistic electromagnetic interaction gives the correction of the flat background metric $\eta_{\mu\nu}$, thus curving spacetime. The curved metric can be nicely broken down into two parts, the pure correction on the flat spacetime metric and the projection tensor. We find that the curved metric is independent of the representation chosen.   
\end{abstract}


\section{Introduction}
The Dirac equation has laid the foundation of relativistic electrodynamics, where the theory of quantum mechanics and special relativity have unified in a whole framework. However, unifying quantum mechanics and general relativity still faces a lot of obstacles due to the problem of renormlizability, and therefore a theory of quantum gravity is yet to be found \cite{ref1,ref2,ref3,ref4,ref5,ref6}. 

In our previous paper, we have demonstrated how to redefine the metric in general relativity using the effective Dirac algebra we novely developed, which enables us to bridge the two theories when the electromagnetic interaction is turned on. The curved metric is expressed in terms of the gauge field $A_{\mu}$ \cite{BWong}. Numerous studies on effective spacetime geometry have been performed in the literature \cite{ext1,ext2,ext3,ext4,ext5,ext6,ext7,ext8,ext9}, while our method provides a novel way to study effective spacetime metric by introducing effective gamma matrices.

We will first recall some essences of the effective Dirac algebra. Starting from the well-known Dirac equation with electromagnetic interaction \cite{DiracAntiMatter,Dirac,Dirac3,Peskin},
\begin{equation} \label{eq:DiracEq}
i \gamma^{\mu} \partial_{\mu} \psi -e\gamma^{\mu}A_{\mu}\psi - m \psi = 0 \,,
\end{equation}
the Dirac gamma matrices allow one to define the flat metric,
\begin{equation} \label{eq:DiracAlgebra}
\{ \gamma^{\mu}, \gamma^{\nu} \} = 2 \eta^{\mu\nu} \pmb{1} \,.
\end{equation}
We rewrite equation (\ref{eq:DiracEq}) as \cite{BWong}
\begin{equation} \label{eq:2}
\psi = \frac{i\hbar}{mc} \frac{1}{1+ \frac{he}{mc} \gamma^{\rho}A_{\rho}} \gamma^{\mu} \partial_{\mu} \psi =\frac{i\lambda}{2\pi}\, \frac{1}{1+ \frac{\lambda e}{2\pi}\gamma^{\rho}A_{\rho}} \gamma^{\mu}\partial_{\mu}\psi = \frac{i\lambda}{2\pi} f[\slashed{A}] \gamma^{\mu}\partial_{\mu}\psi   \,.
\end{equation}
where $\lambda = \frac{h}{mc}$ is the Compton wavelength and
\begin{equation}
f[\slashed{A}] = \frac{1}{1+ \frac{\lambda e}{2\pi} \slashed{A}}
\end{equation}
with $\slashed{A} = A_{\rho}\gamma^{\rho}$ as the Feynman slash notation.
The effective gamma matrix is defined by \cite{BWong}
\begin{equation}
\Gamma^{\mu}_L = f[\slashed{A}] \gamma^{\mu} \,.
\end{equation}
The effective Dirac algebra is given by \cite{BWong},
\begin{equation} \label{eq:EffectiveDiracAlgebra}
\{ \Gamma^{\mu}_L , \Gamma^{\nu}_L \} = 2 g^{\mu\nu}\pmb{1} = 2 f^2 [\slashed{A}] \eta^{\mu\nu} + \frac{\lambda e}{\pi} f[\slashed{A}] \bigg( \frac{1}{1-\frac{\lambda^2 e^2}{4\pi^2}A^2} \bigg) ( 2\eta^{\mu\nu}\slashed{A}-\gamma^\mu A^\nu - \gamma^\nu A^\mu )\,,
\end{equation}
and therefore the curved metric is defined. It is also noted that $A^2 =A_{\rho}A^{\rho} = A_0^2 - A_1^2 -  A_2^2 - A_3^2$. It is clear that when the gauge field is turned off, the effective Dirac algebra in equation (\ref{eq:EffectiveDiracAlgebra}) restores back to the original Dirac algebra in equation (\ref{eq:DiracAlgebra}). It follows that the metric tensor is then given by \cite{BWong}
\begin{equation} \label{eq:metricc}
g_{\mu\nu} = \frac{1}{4}\mathrm{Tr}\, \bigg\{ f^2 [\slashed{A}] \eta_{\mu\nu} + \frac{\lambda e}{2\pi} f[\slashed{A}] \bigg( \frac{1}{1-\frac{\lambda^2 e^2}{4\pi^2}A^2} \bigg) ( 2\eta_{\mu\nu}\slashed{A}-\gamma_\mu A_\nu - \gamma_\nu A_\mu )\bigg\} \,.
\end{equation}
In our previous paper, we have studied the special case of (\ref{eq:metricc}), which is the relativistic hydrogen atom case. The $A^{\mu}$ vector is taken to be
\begin{equation}
    A^{\mu} = 
    \begin{pmatrix}
     -\frac{Ze}{4\pi \epsilon_0 r}\\
     0 \\
     0\\
     0\\
    \end{pmatrix} \,,
\end{equation}
with $Z$ the atomic number and $Z=1$ for the hydrogen case. Using the metric equation in (\ref{eq:metricc}), the final metric is found to be \cite{BWong},
\begin{equation} \label{eq:fullmetric}
\begin{aligned}
   & ds^2 = \bigg[   \frac{1}{2}\bigg( \frac{1}{1-\frac{Ze^2 h}{4\pi^2 m c \epsilon_0 r}} \bigg)^2 + \frac{1}{2}\bigg( \frac{1}{1+\frac{Ze^2 h}{4\pi^2 m c \epsilon_0 r}} \bigg)^2 \bigg] dt^2 \\
   &- \bigg[\frac{1}{2} \bigg( \Big(\frac{1}{1-\frac{Ze^2 h}{4\pi^2 m c \epsilon_0 r}}\Big)^2 +   \Big(\frac{1}{1+\frac{Ze^2 h}{4\pi^2 m c \epsilon_0 r}}\Big)^2  \bigg) +\frac{\frac{Ze^2 h}{4\pi^2 m c \epsilon_0 r}}{1-\Big(\frac{Ze^2 h}{4\pi^2 m c \epsilon_0 r}\Big)^2}\bigg(\frac{1}{1-\frac{Ze^2 h}{4\pi^2 m c \epsilon_0 r}} - \frac{1}{1+\frac{Ze^2 h}{4\pi^2 m c \epsilon_0 r}} \bigg) \bigg] \\
   & \times (dr^2 + r^2 d\theta^2 + r^2 \sin^2 \theta d\phi^2 ) \,.
\end{aligned}
\end{equation}
However, the explicit form of (\ref{eq:metricc}) has not been calculated.

\section{The complete metric by effective Dirac Algebra in Dirac representation }
\label{sec:1}
In this paper, we will compute the explicit metric given by (\ref{eq:metricc}). For convenience, we will break down the metric $g_{\mu\nu}$ in (\ref{eq:metricc}) into two terms,
\begin{equation}
g^{(1)}_{\mu\nu} = \frac{1}{4} \mathrm{Tr}\, \bigg( f^2 [\slashed{A}] \eta_{\mu\nu} \bigg) \,,
\end{equation}
and
\begin{equation}
g^{(2)}_{\mu\nu} = \frac{1}{4} \mathrm{Tr}\, \bigg\{ \frac{\lambda e}{2\pi} f[\slashed{A}] \bigg( \frac{1}{1-\frac{\lambda^2 e^2}{4\pi^2}A^2} \bigg) ( 2\eta_{\mu\nu}\slashed{A}-\gamma_\mu A_\nu - \gamma_\nu A_\mu )  \bigg\} \,.
\end{equation}
Consider the Dirac representation of the $\gamma^{\mu}$ matrix,
\begin{equation}
\gamma^0 =
\begin{pmatrix}
 I & 0 \\
 0 & -I
\end{pmatrix} \,\,,\,\,
\gamma^i =
\begin{pmatrix}
 0 & \sigma^i \\
 -\sigma^i & 0
\end{pmatrix} \,,
\end{equation}
where $I$ is the $2\times 2$ identity matrix and $\sigma^i$ is the three Pauli matrices. Explicitly, we have
\begin{equation}
    \gamma^0 =
    \begin{pmatrix}
    1 & 0 & 0 & 0 \\
    0 & 1 & 0 & 0 \\
    0 & 0 & -1 & 0 \\
    0 & 0 & 0 & -1
    \end{pmatrix}\,,\,
    \gamma^1 =
    \begin{pmatrix}
    0 & 0 & 0 & 1 \\
    0 & 0 & 1 & 0 \\
    0 & -1 & 0 & 0 \\
    -1 & 0 & 0 & 0
    \end{pmatrix}\,,\,
    \gamma^2 =
    \begin{pmatrix}
    0 & 0 & 0 & -i \\
    0 & 0 & i & 0 \\
    0 & i & 0 & 0 \\
    -i & 0 & 0 & 0
    \end{pmatrix}\,,\,
    \gamma^3 =
    \begin{pmatrix}
    0 & 0 & 1 & 0 \\
    0 & 0 & 0 & -1 \\
    -1 & 0 & 0 & 0 \\
    0 & 1 & 0 & 0
    \end{pmatrix},
\end{equation}
First we compute the following term under Dirac representation,
\begin{equation} \label{eq:step1}
\slashed{A} = A_{\rho}\gamma^{\rho} =
\begin{pmatrix}
    A_0 & 0 & A_3 & A_1 - iA_2 \\
    0 & A_0 & A_1 + iA_2 & -A_3 \\
    -A_3 & -A_1 + iA_2 & -A_0 & 0 \\
    -A_1 -iA_2 & A_3 & 0 & -A_0
    \end{pmatrix} \,.
\end{equation}
Then by some tedious algebra we have the inverse matrix as
\begin{equation} \label{eq:Inverse}
\begin{aligned}
f[\slashed{A}] &= \bigg(1+ \frac{\lambda e}{2\pi} A_{\rho}\gamma^{\rho}  \bigg)^{-1} \\
&= \frac{1}{1-\left(\frac{\lambda e}{2\pi}\right)^2 A^2} 
\begin{pmatrix}
    1-\frac{\lambda e}{2\pi} A_0 & 0 & -\frac{\lambda e}{2\pi} A_3 & - \frac{\lambda e}{2\pi} (A_1 - iA_2 ) \\
    0 & 1- \frac{\lambda e}{2\pi} A_0 & -\frac{\lambda e}{2\pi}(A_1 + iA_2) & \frac{\lambda e}{2\pi} A_3 \\
    \frac{\lambda e}{2\pi} A_3 & \frac{\lambda e}{2\pi}(A_1 - iA_2) & 1+ \frac{\lambda e}{2\pi} A_0 & 0 \\
    \frac{\lambda e}{2\pi}(A_1 +iA_2) & -\frac{\lambda e}{2\pi} A_3 & 0 & 1+ \frac{\lambda e}{2\pi} A_0
\end{pmatrix} \,.
\end{aligned}
\end{equation}
Then we have
\begin{equation}
g_{\mu\nu}^{(1)} =\frac{1}{4} \eta_{\mu\nu} \mathrm{Tr}\,\bigg[\bigg(1+ \frac{\lambda e}{2\pi} A_{\rho}\gamma^{\rho}  \bigg)^{-1} \bigg]^2 = \frac{1+ \left( \frac{\lambda e}{2\pi} \right)^2 A^2}{\left( 1-\left( \frac{\lambda e}{2\pi}\right)^2 A^2 \right)^2} \eta_{\mu\nu} \,.
\end{equation}
Now we begin to find out each component of the metric. First as $\eta_{00} = 1$, we have
\begin{equation}
g_{00}^{(1)} =\frac{1+ \left( \frac{\lambda e}{2\pi} \right)^2 A^2}{\left( 1-\left( \frac{\lambda e}{2\pi}\right)^2 A^2 \right)^2} \,.
\end{equation}
Next we need to find $g^{(2)}_{00}$.  Using the results in (\ref{eq:step1}) and(\ref{eq:Inverse}), with some lines of algebra, we obtain
\begin{equation}
\begin{aligned}
g^{(2)}_{00} &= \frac{1}{4} \mathrm{Tr}\, \bigg\{ \frac{\lambda e}{2\pi}  \bigg( \frac{1}{1-\frac{\lambda^2 e^2}{4\pi^2}A^2} \bigg)\bigg(1+ \frac{\lambda e}{2\pi} A_{\rho}\gamma^{\rho}  \bigg)^{-1} ( 2\eta_{00}A_{\sigma}\gamma^\sigma -2\gamma_0 A_0  )  \bigg\} \\
&= \frac{1}{4} \mathrm{Tr}\, \bigg\{ \frac{\lambda e}{2\pi}  \bigg( \frac{1}{1-\frac{\lambda^2 e^2}{4\pi^2}A^2} \bigg)\bigg(1+ \frac{\lambda e}{2\pi} A_{\rho}\gamma^{\rho}  \bigg)^{-1} ( 2 A_{\sigma}\gamma^\sigma -2\gamma^0 A_0  )  \bigg\} \\
&= \frac{-2 \left( \frac{\lambda e}{2\pi}\right)^2 ( A^2 - A_0^2 )}{\left( 1-\left( \frac{\lambda e}{2\pi}\right)^2 A^2 \right)^2} \,.
\end{aligned}
\end{equation}
Therefore,
\begin{equation}
g_{00}=g^{(1)}_{00} + g^{(2)}_{00} =\frac{1+ \left( \frac{\lambda e}{2\pi} \right)^2 A^2}{\left( 1-\left( \frac{\lambda e}{2\pi}\right)^2 A^2 \right)^2} + \frac{-2 \left( \frac{\lambda e}{2\pi}\right)^2 ( A^2 - A_0^2 )}{\left( 1-\left( \frac{\lambda e}{2\pi}\right)^2 A^2 \right)^2} = \frac{1- \left( \frac{\lambda e}{2\pi}\right)^2 (A^2 -2A_0^2)}{\left( 1-\left( \frac{\lambda e}{2\pi}\right)^2 A^2 \right)^2} \,.
\end{equation}
Next, we find the component $g_{ii}$. First, as $\eta_{ii} = -1$, 
\begin{equation}
g_{ii}^{(1)} =- \frac{1+ \left( \frac{\lambda e}{2\pi} \right)^2 A^2}{\left( 1-\left( \frac{\lambda e}{2\pi}\right)^2 A^2 \right)^2} \,.
\end{equation}
Then we need to compute $g_{11}$, $g_{22}$ and $g_{33}$ independently due to different $\gamma^1$, $\gamma^2$, $\gamma^3$ matrices. As $\eta_{11}=\eta_{22}=\eta_{33} = -1$, we  have $g_{11}^{(1)} =g_{22}^{(1)} = g_{33}^{(1)} $.
Now we compute $g^{(2)}_{11}$, $g^{(2)}_{22}$ and $g^{(2)}_{33}$.
\begin{equation}
\begin{aligned}
g^{(2)}_{11} &= \frac{1}{4} \mathrm{Tr}\, \bigg\{ \frac{\lambda e}{2\pi}  \bigg( \frac{1}{1-\frac{\lambda^2 e^2}{4\pi^2}A^2} \bigg)\bigg(1+ \frac{\lambda e}{2\pi} A_{\rho}\gamma^{\rho}  \bigg)^{-1} ( 2\eta_{11}A_{\sigma}\gamma^\sigma -2\gamma_1 A_1  )  \bigg\} \\
&= \frac{1}{4} \mathrm{Tr}\, \bigg\{ \frac{\lambda e}{2\pi}  \bigg( \frac{1}{1-\frac{\lambda^2 e^2}{4\pi^2}A^2} \bigg)\bigg(1+ \frac{\lambda e}{2\pi} A_{\rho}\gamma^{\rho}  \bigg)^{-1} ( -2 A_{\sigma}\gamma^\sigma + 2\gamma^1 A_1 )  \bigg\} \\
&= \frac{2 \left( \frac{\lambda e}{2\pi}\right)^2 ( A^2 + A_1^2 )}{\left( 1-\left( \frac{\lambda e}{2\pi}\right)^2 A^2 \right)^2} \,.
\end{aligned}
\end{equation}
Similarly, one can find that
\begin{equation}
\begin{aligned}
g^{(2)}_{22} &= \frac{1}{4} \mathrm{Tr}\, \bigg\{ \frac{\lambda e}{2\pi}  \bigg( \frac{1}{1-\frac{\lambda^2 e^2}{4\pi^2}A^2} \bigg)\bigg(1+ \frac{\lambda e}{2\pi} A_{\rho}\gamma^{\rho}  \bigg)^{-1} ( 2\eta_{22}A_{\sigma}\gamma^\sigma -2\gamma_2 A_2  )  \bigg\} \\
&= \frac{1}{4} \mathrm{Tr}\, \bigg\{ \frac{\lambda e}{2\pi}  \bigg( \frac{1}{1-\frac{\lambda^2 e^2}{4\pi^2}A^2} \bigg)\bigg(1+ \frac{\lambda e}{2\pi} A_{\rho}\gamma^{\rho}  \bigg)^{-1} ( -2 A_{\sigma}\gamma^\sigma + 2\gamma^2 A_2 )  \bigg\} \\
&= \frac{2 \left( \frac{\lambda e}{2\pi}\right)^2 ( A^2 + A_2^2 )}{\left( 1-\left( \frac{\lambda e}{2\pi}\right)^2 A^2 \right)^2} \,.
\end{aligned}
\end{equation}
And finally,
\begin{equation}
\begin{aligned}
g^{(2)}_{33} &= \frac{1}{4} \mathrm{Tr}\, \bigg\{ \frac{\lambda e}{2\pi}  \bigg( \frac{1}{1-\frac{\lambda^2 e^2}{4\pi^2}A^2} \bigg)\bigg(1+ \frac{\lambda e}{2\pi} A_{\rho}\gamma^{\rho}  \bigg)^{-1} ( 2\eta_{33}A_{\sigma}\gamma^\sigma -2\gamma_3 A_3  )  \bigg\} \\
&= \frac{1}{4} \mathrm{Tr}\, \bigg\{ \frac{\lambda e}{2\pi}  \bigg( \frac{1}{1-\frac{\lambda^2 e^2}{4\pi^2}A^2} \bigg)\bigg(1+ \frac{\lambda e}{2\pi} A_{\rho}\gamma^{\rho}  \bigg)^{-1} ( -2 A_{\sigma}\gamma^\sigma + 2\gamma^3 A_3 )  \bigg\} \\
&= \frac{2 \left( \frac{\lambda e}{2\pi}\right)^2 ( A^2 + A_3^2 )}{\left( 1-\left( \frac{\lambda e}{2\pi}\right)^2 A^2 \right)^2} \,.
\end{aligned}
\end{equation}
Therefore generally, 
\begin{equation}
g_{ii}^{(2)} =  \frac{2 \left( \frac{\lambda e}{2\pi}\right)^2 ( A^2 + A_i^2 )}{\left( 1-\left( \frac{\lambda e}{2\pi}\right)^2 A^2 \right)^2} \,,
\end{equation}
and hence
\begin{equation}
g_{ii} = g_{ii}^{(1)} + g_{ii}^{(2)} = - \frac{1+ \left( \frac{\lambda e}{2\pi} \right)^2 A^2}{\left( 1-\left( \frac{\lambda e}{2\pi}\right)^2 A^2 \right)^2} + \frac{2 \left( \frac{\lambda e}{2\pi}\right)^2 ( A^2 + A_i^2 )}{\left( 1-\left( \frac{\lambda e}{2\pi}\right)^2 A^2 \right)^2} = -\frac{1- \left( \frac{\lambda e}{2\pi}\right)^2 (A^2 +2A_i^2)}{\left( 1-\left( \frac{\lambda e}{2\pi}\right)^2 A^2 \right)^2} \,.
\end{equation}
Next we find the $g_{0i}$ components. Since $\eta_{0i} = 0$, the first term $g_{0i}^{(1)}$ vanishes. The remaining non-vanishing terms are,
\begin{equation} \label{eq:0i}
g_{0i}^{(2)} = \frac{1}{4} \mathrm{Tr}\, \bigg\{ \frac{\lambda e}{2\pi}  \bigg( \frac{1}{1-\frac{\lambda^2 e^2}{4\pi^2}A^2} \bigg)\bigg(1+ \frac{\lambda e}{2\pi} A_{\rho}\gamma^{\rho}  \bigg)^{-1} ( -\gamma_0 A_i -\gamma_i A_0 )  \bigg\} \,. 
\end{equation}
The last two matrix terms are non-vanishing, giving us the following,
\begin{equation}
-\gamma_0 A_1 -\gamma_1 A_0 = -\gamma^0 A_1 + \gamma^1 A_0 =
\begin{pmatrix}
-A_1 & 0 & 0 & A_0 \\
0 & -A_1 & A_0 & 0 \\
0 & -A_0 & A_1 & 0 \\
-A_0 & 0 & 0 & A_1 
\end{pmatrix}  \,\,,
\end{equation}
and
\begin{equation}
-\gamma_0 A_2 -\gamma_2 A_0 = -\gamma^0 A_2 + \gamma^2 A_0 =
\begin{pmatrix}
-A_2 & 0 & 0 & -iA_0 \\
0 & -A_2 & iA_0 & 0 \\
0 & iA_0 & A_2 & 0 \\
-iA_0 & 0 & 0 & A_2 
\end{pmatrix} \,\,,
\end{equation}
and
\begin{equation}
-\gamma_0 A_3 -\gamma_3 A_0 = -\gamma^0 A_3 + \gamma^3 A_0 =
\begin{pmatrix}
-A_3 & 0 & A_0 & 0 \\
0 & -A_3 & 0 & -A_0 \\
-A_0 & 0 & A_3 & 0 \\
0 & A_0 & 0 & A_3 
\end{pmatrix} \,\,.
\end{equation}
Substituting these results back into equation (\ref{eq:0i}), after some tedious algebra, we obtain the following results,
\begin{equation}
g_{01} = \frac{2 \left(\frac{\lambda e}{2\pi}\right)^2 A_0 A_1}{\left( 1-\left( \frac{\lambda e}{2\pi}\right)^2 A^2 \right)^2} \,\, , \,\, g_{02} = \frac{2 \left(\frac{\lambda e}{2\pi}\right)^2 A_0 A_2}{\left( 1-\left( \frac{\lambda e}{2\pi}\right)^2 A^2 \right)^2} \,\,, \,\, g_{03}  = \frac{2 \left(\frac{\lambda e}{2\pi}\right)^2 A_0 A_3}{\left( 1-\left( \frac{\lambda e}{2\pi}\right)^2 A^2 \right)^2} \,.
\end{equation} 
Generally,
\begin{equation}
g_{0i} = \frac{2 \left(\frac{\lambda e}{2\pi}\right)^2 A_0 A_i}{\left( 1-\left( \frac{\lambda e}{2\pi}\right)^2 A^2 \right)^2} \,.
\end{equation}
Finally, we have to compute the $g_{ij}$ elements. There are three of them we need to calculate, namely $g_{12}$, $g_{13}$ and $g_{23}$. Since $\eta_{ij}=0$, we have $g_{ij}^{(1)}=0$. The non-vanishing terms are
\begin{equation} \label{eq:ij}
g_{ij}^{(2)} = \frac{1}{4} \mathrm{Tr}\, \bigg\{ \frac{\lambda e}{2\pi}  \bigg( \frac{1}{1-\frac{\lambda^2 e^2}{4\pi^2}A^2} \bigg)\bigg(1+ \frac{\lambda e}{2\pi} A_{\rho}\gamma^{\rho}  \bigg)^{-1} ( -\gamma_i A_j -\gamma_i A_j )  \bigg\} \,. 
\end{equation}
We evaluate the last two terms of matrix,
\begin{equation}
-\gamma_1 A_2 -\gamma_2 A_1 = \gamma^1 A_2 +\gamma^2 A_1 =
\begin{pmatrix}
0 & 0 & 0 & A_2 -iA_1 \\
0 & 0 & A_2 +iA_1 &  0 \\
0 &-A_2 + iA_1 & 0 & 0 \\
-A_2 -iA_1 & 0 & 0 & 0 
\end{pmatrix}
\end{equation}
and
\begin{equation}
-\gamma_1 A_3 -\gamma_3 A_1 = \gamma^1 A_3 +\gamma^3 A_1 =
\begin{pmatrix}
0 & 0 & A_1 & A_3 \\
0 & 0 & A_3 & -A_1 \\
-A_1 & -A_3 & 0 & 0 \\
-A_3 & A_1 & 0 & 0 
\end{pmatrix}
\end{equation}
and
\begin{equation}
-\gamma_2 A_3 -\gamma_3 A_2 = \gamma^2 A_3 +\gamma^3 A_2 =
\begin{pmatrix}
0 & 0 & A_2 & -iA_3 \\
0 & 0 & iA_3 & -A_2 \\
-A_2 & iA_3 & 0 & 0 \\
iA_3 & -A_2 & 0 & 0 
\end{pmatrix}
\end{equation}
Substituting these results back into equation (\ref{eq:ij}), we obtain the following results,
\begin{equation}
g_{12} = \frac{2 \left(\frac{\lambda e}{2\pi}\right)^2 A_1 A_2}{\left( 1-\left( \frac{\lambda e}{2\pi}\right)^2 A^2 \right)^2} \,\, , \,\, g_{13} = \frac{2 \left(\frac{\lambda e}{2\pi}\right)^2 A_1 A_3}{\left( 1-\left( \frac{\lambda e}{2\pi}\right)^2 A^2 \right)^2} \,\,, \,\, g_{23}  = \frac{2 \left(\frac{\lambda e}{2\pi}\right)^2 A_2 A_3}{\left( 1-\left( \frac{\lambda e}{2\pi}\right)^2 A^2 \right)^2} \,.
\end{equation} 
Generally
\begin{equation}
g_{ij} = \frac{2 \left(\frac{\lambda e}{2\pi}\right)^2 A_i A_j}{\left( 1-\left( \frac{\lambda e}{2\pi}\right)^2 A^2 \right)^2} \,.
\end{equation}
Hence we have completed our calculation of all the independent components of the metric tensor, as the remaining components can be found by the fact that the metric is symmetric $g_{\mu\nu} = g_{\nu\mu}$.  The full metric tensor now reads,
\begin{equation} \label{eq:finaltensor}
\begin{small}
\begin{aligned}
&g_{\mu\nu}= \\
&\frac{1}{l^2}
\begin{pmatrix}
1- \left( \frac{\lambda e}{2\pi}\right)^2 (A^2 -2A_0^2) &  \left( \frac{\lambda e}{2\pi}\right)^2 A_0 A_1 & \left( \frac{\lambda e}{2\pi}\right)^2 A_0 A_2 & \left( \frac{\lambda e}{2\pi}\right)^2 A_0 A_3 \\
\left( \frac{\lambda e}{2\pi}\right)^2 A_1 A_0 & -1 + \left( \frac{\lambda e}{2\pi}\right)^2 (A^2 +2A_1^2) & \left( \frac{\lambda e}{2\pi}\right)^2 A_1 A_2 
& \left( \frac{\lambda e}{2\pi}\right)^2 A_1 A_3 \\
\left( \frac{\lambda e}{2\pi}\right)^2 A_2 A_0 & \left( \frac{\lambda e}{2\pi}\right)^2 A_2 A_1 & -1+ \left( \frac{\lambda e}{2\pi}\right)^2 (A^2 +2A_2^2) & \left( \frac{\lambda e}{2\pi}\right)^2 A_2 A_3   \\
\left( \frac{\lambda e}{2\pi}\right)^2 A_3 A_0 & \left( \frac{\lambda e}{2\pi}\right)^2 A_3 A_1&  \left( \frac{\lambda e}{2\pi}\right)^2 A_3 A_2&- 1 + \left( \frac{\lambda e}{2\pi}\right)^2 (A^2 +2A_3^2)
\end{pmatrix} \,,
\end{aligned}
\end{small}
\end{equation}
where $l^2 = \left( 1-\left( \frac{\lambda e}{2\pi}\right)^2 A^2 \right)^2 $ is the normalization. It is important to see that, although the gamma matrix is complex in nature and there are complex matrices during our intermediate steps of calculation, the final metric must be complex-number free. Finally, we can easily check that, when the interaction is turned off, i.e. $A_{\mu} =0$, the above metric tensor reduces back to the Minkowski metric. 

Upon detailed speculation, we can see that (\ref{eq:finaltensor}) can be beautifully separated into two parts,
\begin{equation} \label{eq:metric2}
\begin{aligned}
g_{\mu\nu} &=\frac{1}{l^2}
\begin{pmatrix}
1+\left(\frac{\lambda e}{2\pi}\right)^2 A^2 & 0 & 0 & 0 \\
0 & -\left(1+\left(\frac{\lambda e}{2\pi}\right)^2 A^2 \right) & 0 & 0 \\
0 & 0 & -\left(1+\left(\frac{\lambda e}{2\pi}\right)^2 A^2 \right) & 0 \\
0 & 0 & 0 & -\left(1+\left(\frac{\lambda e}{2\pi}\right)^2 A^2 \right)
\end{pmatrix} \\
&-\frac{2}{l^2} \left(\frac{\lambda e}{2\pi}\right)^2
\begin{pmatrix}
A^2 -A_0^2 & -A_0 A_1 & -A_0 A_2 & -A_0 A_3 \\
-A_1 A_0 & -(A^2 +A_1^2) & -A_1 A_2 & -A_1 A_3 \\
-A_2 A_0 & -A_2 A_1 & -(A^2 +A_2^2 ) & -A_2 A_3 \\
-A_3 A_0 & -A_3 A_1 & -A_3 A_2 &- (A^2 +A_3^2)
\end{pmatrix}
\end{aligned}
\end{equation}
The first matrix in (\ref{eq:metric2}) is considered as the correction of the flat background metric $\eta_{\mu\nu}$, which is given by $g_{\mu\nu}^{(1)}$. The second matrix in  (\ref{eq:metric2}) is the normalized projection tensor $P_{\mu\nu}$, which is given by $g_{\mu\nu}^{(2)}$. Therefore,
\begin{equation}
\begin{aligned}
g^{(2)}_{\mu\nu} &= \frac{1}{4} \mathrm{Tr}\, \bigg\{ \frac{\lambda e}{2\pi} f[\slashed{A}] \bigg( \frac{1}{1-\frac{\lambda^2 e^2}{4\pi^2}A^2} \bigg) ( 2\eta_{\mu\nu}\slashed{A}-\gamma_\mu A_\nu - \gamma_\nu A_\mu )  \bigg\} =P_{\mu\nu} \\
&=-\frac{2}{ \left( 1-\left( \frac{\lambda e}{2\pi}\right)^2 A^2 \right)^2} \left(\frac{\lambda e}{2\pi}\right)^2 (A^2 \eta_{\mu\nu}- A_{\mu}A_{\nu}) \,. 
\end{aligned}
\end{equation}
Therefore, the final full metric is
\begin{equation} \label{eq:finalmetric}
g_{\mu\nu} = \frac{1}{ \left( 1-\left( \frac{\lambda e}{2\pi}\right)^2 A^2 \right)^2} \bigg[ \Big( 1+ \left( \frac{\lambda e}{2\pi}\right)^2 A^2 \Big)\eta_{\mu\nu} -2 \left( \frac{\lambda e}{2\pi}\right)^2 (A^2 \eta_{\mu\nu}- A_{\mu}A_{\nu})  \bigg] \,.
\end{equation}
Equation (\ref{eq:finalmetric}) is the final complete metric  we obtain from the effective Dirac Algebra in Dirac representation. 

\section{The complete metric by effective Dirac Algebra in Weyl representation }

Consider the Weyl representation of the $\gamma^{\mu}$ matrices,
\begin{equation}
\gamma^0 =
\begin{pmatrix}
 0 & I \\
 I & 0
\end{pmatrix} \,\,,\,\,
\gamma^i =
\begin{pmatrix}
 0 & \sigma^i \\
 -\sigma^i & 0
\end{pmatrix} \,.
\end{equation}
The only difference between the Dirac representation and Weyl representation is the $\gamma^0$ matrix, while it is the same for the remaining $\gamma^i$ matrices. Explicitly,
\begin{equation}
    \gamma^0 =
    \begin{pmatrix}
    0 & 0 & 1 & 0 \\
    0 & 0 & 0 & 1 \\
    1 & 0 & 0 & 0 \\
    0 & 1 & 0 & 0
    \end{pmatrix}\,,\,
    \gamma^1 =
    \begin{pmatrix}
    0 & 0 & 0 & 1 \\
    0 & 0 & 1 & 0 \\
    0 & -1 & 0 & 0 \\
    -1 & 0 & 0 & 0
    \end{pmatrix}\,,\,
    \gamma^2 =
    \begin{pmatrix}
    0 & 0 & 0 & -i \\
    0 & 0 & i & 0 \\
    0 & i & 0 & 0 \\
    -i & 0 & 0 & 0
    \end{pmatrix}\,,\,
    \gamma^3 =
    \begin{pmatrix}
    0 & 0 & 1 & 0 \\
    0 & 0 & 0 & -1 \\
    -1 & 0 & 0 & 0 \\
    0 & 1 & 0 & 0
    \end{pmatrix} \,.
\end{equation}
Then we have
\begin{equation} \label{eq:step1}
\slashed{A} = A_{\rho}\gamma^{\rho} =
\begin{pmatrix}
    0 & 0 &  A_0 + A_3 & A_1 - iA_2 \\
    0 & 0 & A_1 + iA_2 & A_0 -A_3 \\
    A_0 -A_3 & -A_1 + iA_2 & 0 & 0 \\
    -A_1 -iA_2 & A_0 + A_3 & 0 & 0
    \end{pmatrix} \,.
\end{equation}
Then by some tedious algebra we have
\begin{equation} \label{eq:Inverse}
\begin{aligned}
f[\slashed{A}] &= \bigg(1+ \frac{\lambda e}{2\pi} A_{\rho}\gamma^{\rho}  \bigg)^{-1} \\
&= \frac{1}{1-\left(\frac{\lambda e}{2\pi}\right)^2 A^2} 
\begin{pmatrix}
    1 & 0 & -\frac{\lambda e}{2\pi} (A_0 + A_3) & - \frac{\lambda e}{2\pi} (A_1 - iA_2 ) \\
    0 & 1 & -\frac{\lambda e}{2\pi}(A_1 + iA_2) & -\frac{\lambda e}{2\pi} (A_0 -A_3) \\
    -\frac{\lambda e}{2\pi}(A_0- A_3) & \frac{\lambda e}{2\pi}(A_1 - iA_2) & 1 & 0 \\
    \frac{\lambda e}{2\pi}(A_1 +iA_2) & -\frac{\lambda e}{2\pi} (A_0 + A_3 ) & 0 & 1 
\end{pmatrix} \,.
\end{aligned}
\end{equation}
Then we compute 
\begin{equation}
g_{\mu\nu}^{(1)} =\frac{1}{4} \eta_{\mu\nu} \mathrm{Tr}\,\bigg[\bigg(1+ \frac{\lambda e}{2\pi} A_{\rho}\gamma^{\rho}  \bigg)^{-1} \bigg]^2 = \frac{1+ \left( \frac{\lambda e}{2\pi} \right)^2 A^2}{\left( 1-\left( \frac{\lambda e}{2\pi}\right)^2 A^2 \right)^2} \eta_{\mu\nu} \,,
\end{equation}
which is same as the case as the Dirac representation. And one finds,
\begin{equation}
g_{00}^{(2)} = \frac{-2 \left( \frac{\lambda e}{2\pi}\right)^2 ( A^2 - A_0^2 )}{\left( 1-\left( \frac{\lambda e}{2\pi}\right)^2 A^2 \right)^2} \,,
\end{equation}
which is again same as the Dirac representation.
Therefore, the $g_{00}$ result computed in the Weyl representation is same as the one computed in Dirac representation.

Next, the $g_{ii}$ components will be the same as the Dirac representation because only the $\gamma^0$ component is different.

Now we compute the $g_{0i}$ components. Again we compute the three matrices,
\begin{equation}
-\gamma_0 A_1 -\gamma_1 A_0 = -\gamma^0 A_1 + \gamma^1 A_0 =
\begin{pmatrix}
0 & 0 & -A_1 & A_0 \\
0 & 0 & A_0 & -A_1 \\
-A_1 & -A_0 & 0 & 0 \\
-A_0 & -A_1 & 0 & 0 
\end{pmatrix}  \,\,,
\end{equation}
and
\begin{equation}
-\gamma_0 A_2 -\gamma_2 A_0 = -\gamma^0 A_2 + \gamma^2 A_0 =
\begin{pmatrix}
0 & 0 & -A_2 & -iA_0 \\
0 & 0 & iA_0 & -A_2 \\
-A_2 & iA_0 & 0 & 0 \\
-iA_0 & -A_2 & 0 & 0 
\end{pmatrix} \,\,,
\end{equation}
and
\begin{equation}
-\gamma_0 A_3 -\gamma_3 A_0 = -\gamma^0 A_3 + \gamma^3 A_0 =
\begin{pmatrix}
0 & 0 & -A_3+A_0 & 0 \\
0 & 0 & 0 & -A_3-A_0 \\
-A_3 -A_0 & 0 & A_3 & 0 \\
0 & -A_3 +A_0 & 0 & A_3 
\end{pmatrix} \,\,.
\end{equation}
Then we find that,
\begin{equation}
g_{01} = \frac{2 \left(\frac{\lambda e}{2\pi}\right)^2 A_0 A_1}{\left( 1-\left( \frac{\lambda e}{2\pi}\right)^2 A^2 \right)^2} \,\, , \,\, g_{02} = \frac{2 \left(\frac{\lambda e}{2\pi}\right)^2 A_0 A_2}{\left( 1-\left( \frac{\lambda e}{2\pi}\right)^2 A^2 \right)^2} \,\,, \,\, g_{03}  = \frac{2 \left(\frac{\lambda e}{2\pi}\right)^2 A_0 A_3}{\left( 1-\left( \frac{\lambda e}{2\pi}\right)^2 A^2 \right)^2} \,,
\end{equation} 
which is same as the case as the Dirac representation, even though the matrix representations are different.  

For $g_{ij}$, the case by Weyl representation would be same as the case by Dirac representation as the $\gamma^i$ remains the same for both representations. 

Therefore, the full metric remains the same as (\ref{eq:finalmetric}) for the Weyl representation. This demonstrates that the metric is independent of the representations of the Dirac gamma matrices.

\section{Conclusion}
In this article, we have presented the complete study of the curved metric by the effective Dirac algebra under Dirac and Weyl representation. We find that the metric is independent of the representation chosen. The effect of electromagnetic gauge field comes into following as the correction of Minkowski flat metric background and the projection tensor. When the interaction turns off, this gives us back the original Minkowski metric.

\section{Declaration}
I declare that there are no known competing financial interests or personal relationships that could have appeared to influence the work reported in this paper.

\end{document}